\documentclass{PoS}

\usepackage{amsmath}
\usepackage{amssymb}
\usepackage{color}
\usepackage{collref}

\newcommand{\gev}{~\mathrm{GeV}}
\newcommand{\tev}{~\mathrm{TeV}}
\newcommand{\pb}{~\mathrm{pb}}
\newcommand{\fb}{~\mathrm{fb}}
\newcommand{\HB}{\texttt{HiggsBounds}}
\newcommand{\HS}{\texttt{HiggsSignals}}

\title{Higgs physics beyond the Standard Model}

\ShortTitle{BSM Higgs physics}

\author{\speaker{Tim Stefaniak}\\ 
        Deutsches Elektronen-Synchrotron DESY\\
        E-mail: \email{tim.stefaniak@desy.de}}

\abstract{We give a brief review of beyond-the-Standard Model (BSM) extensions of the Standard Model (SM) Higgs sector. Going from very simple to more complicated models, our survey includes models with additional scalar singlet fields, the Two Higgs Doublet Model (2HDM) and the Higgs sector of the Minimal Supersymmetric Standard Model (MSSM).  We discuss the impact of current experimental results from LHC Higgs searches and measurements and the prospective reach of the LHC in the high-luminosity (HL) phase. We furthermore highlight possible new collider signatures within these models that have not been experimentally probed to date.}

\FullConference{ALPS 2019 An Alpine LHC Physics Summit \\
		April 22 - 27, 2019\\
		Obergurgl, Austria}

\dedicated{DESY 19-150}

\begin{document}

\section{Introduction}
\label{sec:intro}

In the Standard Model (SM) of particle physics one fundamental complex scalar $SU(2)_L$ doublet field $\Phi$ with hypercharge $+\tfrac{1}{2}$ and a scalar potential $V(\Phi) = \mu^2 \Phi^\dagger \Phi + \lambda (\Phi^\dagger \Phi)^2$ is responsible for the spontaneous breaking of the electroweak symmetry. This Brout-Englert-Higgs (BEH) mechanism successfully explains the non-zero masses of the electroweak ($W^\pm, Z$) gauge bosons, as well as the SM fermion masses via gauge-invariant and renormalizable interactions, restores unitarity of the scattering amplitude, and predicts the existence of one fundamental scalar particle --- the Higgs boson.  The discovery of a scalar boson with a mass $M \simeq 125\gev$ at the LHC in 2012 and the on-going measurements of its properties thus far confirm this SM picture --- within the current experimental precision. The discovered scalar particle is a ``SM-like'' Higgs boson. On the other hand, the shape of the scalar potential $V(\Phi)$ is yet to be confirmed experimentally.\footnote{An independent determination of the quartic interaction $\lambda$ requires the measurement of the Higgs self-coupling which will not be possible at the LHC to good-enough precision~\cite{Cepeda:2019klc}.}

At first glance, the minimality of the scalar sector of the SM as well as its effectiveness in describing all current experimental data are quite convincing that we have finally completed the particle physics picture. At second glance, however, we find that the Higgs boson, or more generally, the scalar potential, is a unique place to anticipate effects from new physics beyond the SM (BSM). The Higgs field may interact with the dark matter (DM) sector through a so-called Higgs portal~\cite{Arcadi:2019lka}, and may play a crucial role in the generation of the baryon asymmetry of the Universe~\cite{Morrissey:2012db,Servant:2013uwa}. BSM theories addressing the so-called hierarchy problem, i.e.~the quadratic sensitivity of the Higgs mass parameter to the UV cutoff scale which in turn requires an ``unnatural'' fine-tuning of the bare mass parameter, generally modify or extend the Higgs sector. One of such theories is Supersymmetry (SUSY), and supersymmetric versions of the SM contain at least two Higgs doublets. Other aspects that may motivate an extension of the scalar sector are a possible improvement of the stability of the vacuum or an explanation of (yet-inconclusive) experimental anomalies seen in the current data (e.g., see Ref.~\cite{Wittbrodt:ALPS2019} and Refs.~\cite{Crivellin:2019mvj}, respectively, for discussions on either topic at this workshop).

Quite generally, new physics in the scalar sector can lead to three observational effects: (\emph{i})~modifications of the $125\gev$ Higgs boson properties (couplings, decay rates, $CP$-properties); (\emph{ii}) existence of additional electrically neutral or charged scalar bosons; and (\emph{iii}) interactions of the Higgs boson (and other scalar bosons) with other new particles present in the BSM theory (e.g.~supersymmetric particles). Obviously, the Higgs sector is an exciting place to look for new physics effects, and needs to be studied in detail at present and future colliders.

So far, experimental results from Run~I and Run~II of the LHC have been rather disillusioning. All measurements of the $125\gev$ Higgs boson properties are (within the current experimental precision) in agreement with the SM predictions, and searches for additional scalar states have not found any convincing hints of new particles. In addition, LHC searches for supersymmetric particles or other exotic new particles, DM direct detection experiments, as well as searches for electric dipole moments, have not found evidence for new physics yet. All these experimental facts lead to important constraints on the new physics landscape. Therefore, in this work, we address the following three questions:
\begin{enumerate}
\setlength\itemsep{-0.2em}
\item What do these (non-)observations tell us about new physics?
\item How much more can we probe in the future (at the LHC)?
\item Have we looked everywhere? Could we have missed a BSM signal?
\end{enumerate}
Regarding the coupling properties of the $125\gev$ Higgs boson, the sensitivity to new physics can be assessed in an (in principle) model-independent way in the framework of an effective field theory, as long as the new physics is too heavy or too weakly coupled to be directly accessible at the experiment (see Ref.~\cite{Mimasu:ALPS2019} for a discussion at this workshop). In contrast, in this work we focus on specific renormalizable BSM models. While such studies are by definition model-dependent, this approach is highly predictive, has no validity restrictions (beyond those inherent to the model) and --- most importantly --- enables us to study the possible complementarity of different observables, e.g.~searches for additional scalar bosons with the Higgs rate measurements, or even farther, with flavor or dark matter observables. In order to confront BSM models with the experimental results from Higgs searches and measurements we largely employ the public computer tools \HB~\cite{Bechtle:2013wla,Bechtle:2015pma} and \HS~\cite{Bechtle:2013xfa,Bechtle:2014ewa}. We discuss scalar singlet extensions of the SM in Sec.~\ref{sec:singlets} and scalar doublet extensions in Sec.~\ref{sec:doublets}. We conclude in Sec.~\ref{sec:conclusions}.

\section{Models with additional scalar singlets}
\label{sec:singlets}

\subsection{Adding a real scalar singlet}
\label{subsec:singlet}

Arguably the simplest extension of the SM Higgs sector is the addition of one real scalar degree of freedom $S$ which is a singlet under the SM gauge group. Assuming a discrete $\mathbb{Z}_2$ symmetry, the scalar potential is given by
\begin{align}
V(\Phi, S) = \mu_\Phi^2 \Phi^\dagger \Phi + \mu_S^2 S^2 + \lambda_1 (\Phi^\dagger \Phi)^2 + \lambda_2 S^4 + \lambda_3 \Phi^\dagger \Phi S^2.
\end{align}
If $S$ does not acquire a vacuum expectation value (vev), $\langle S \rangle = 0$, $S$ is stable and thus a highly constrained DM candidate (see Ref.~\cite{Scott:ALPS2019,Athron:2018hpc} for a discussion at this workshop). No mixing between $S$ and the scalar doublet $\Phi$ occurs. In contrast, if $\langle S \rangle \ne 0$ the two scalar fields $S$ and $\Phi$ mix, forming the two physical scalar states $h_{125}$ and $h_S$ (where $h_{125}$ is identified with the observed Higgs boson), and leading to the following collider signatures:
\begin{itemize}
\setlength\itemsep{-0.2em}
\item[(\emph{i})] A reduced signal strength of the $125\gev$ Higgs boson as the Higgs couplings to SM fermions and gauge bosons are universally suppressed by the mixing angle, $\sin\alpha$;
\item[(\emph{ii})] An additional scalar boson, $h_S$, may be searched for at the LHC, produced and decaying identically as a SM Higgs boson with the same mass, but with a highly reduced signal rate;
\item[(\emph{iii})] If $h_S$ is heavy enough it can decay into two SM-like Higgs bosons, $h_S \to h_{125}h_{125}$;
\item[(\emph{iv})] If $h_S$ is light enough, $h_{125}$ can decay into two light scalar bosons, $h_{125}\to h_S h_S$.
\end{itemize}

\begin{figure}[t]
\centering
\includegraphics[width=0.46\textwidth, trim=0cm 1cm 0cm 1cm]{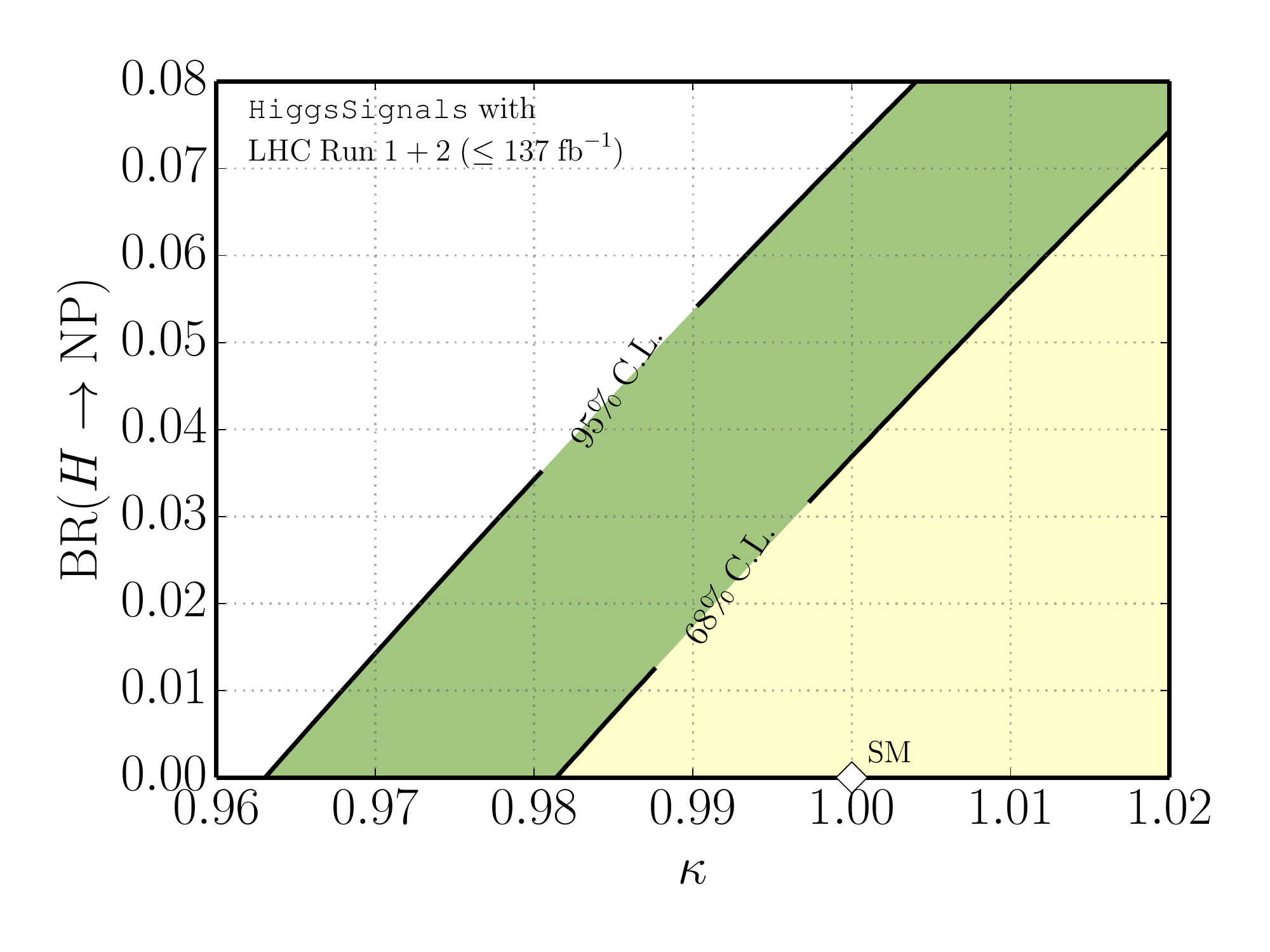}\hfill
\includegraphics[width=0.46\textwidth, trim=0cm 1cm 0cm 1cm]{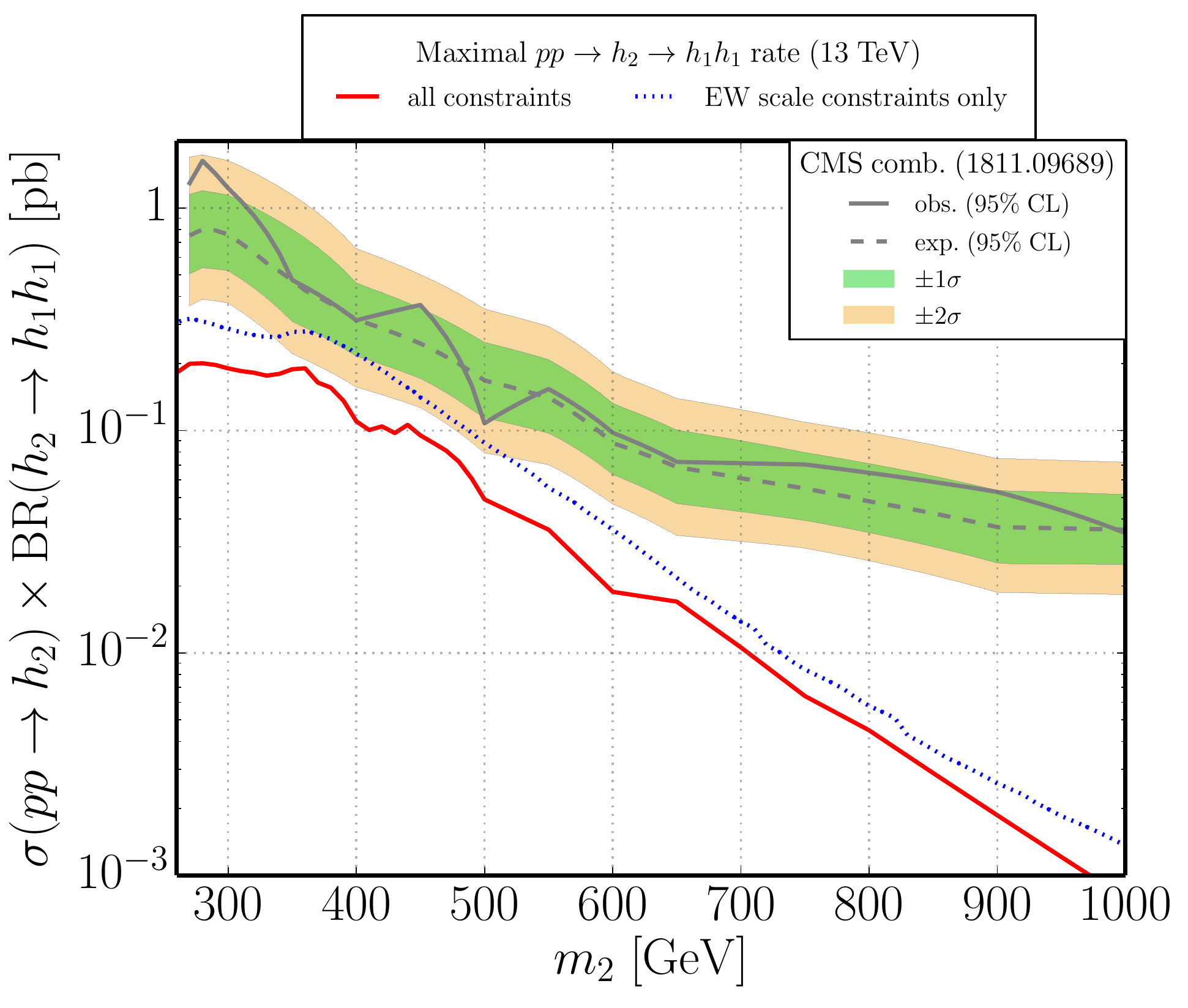}
\caption{\emph{Real scalar singlet extension of the SM:} Constraints on the mixing angle, $\kappa\equiv \sin\alpha$, and the rate of a possible new physics (NP) decay mode, $\mathrm{BR}(H \to\text{NP}) \equiv \mathrm{BR}(h_{125} \to h_S h_S)$, arising from the most recent Higgs rate measurements from ATLAS and CMS (\emph{left panel}); maximal signal rate for $pp \to h_S \to h_{125} h_{125}$ at the $13~\mathrm{TeV}$ LHC, shown as red solid [all constraints applied] and blue dotted line [only EW-scale constraints applied], compared to the current experimental limit from a combination of CMS searches~\cite{Sirunyan:2018two} (\emph{right panel}).}
\label{fig:singlet}
\end{figure}

Detailed phenomenological studies of this model have been presented e.g.~in Refs.~\cite{Robens:2015gla,Robens:2016xkb,Ilnicka:2018def}. The imprint of the first and  fourth signature on the $h_{125}$ properties can be simultaneously described in terms of a universal coupling scale factor $\kappa \equiv \sin\alpha$ and a generic branching ratio (BR) of the Higgs decay to new physics, $\mathrm{BR}(H\to \text{NP}) \equiv \mathrm{BR}(h_{125} \to h_S h_S)$. Taking into account all available Higgs signal rate measurements  from LHC Run~I and Run~II (up to $137~\mathrm{fb}^{-1}$ [\emph{status: July 2019}]) we use \texttt{HiggsSignals} to obtain the $68\%~\mathrm{C.L.}$ and $95\%~\mathrm{C.L.}$ allowed parameter region shown in Fig.~\ref{fig:singlet}~(\emph{left}). These constraints imply that the coupling strength of the light scalar $h_S$ must be significantly reduced with respect to the SM value, $g/g_{SM} = \cos\alpha \lesssim 0.26$ or even less.\footnote{Depending on the mass of $h_S$, further constraints from direct LEP Higgs searches~\cite{Barate:2003sz} may be even stronger~\cite{Robens:2015gla,Robens:2016xkb,Ilnicka:2018def}.} More generally, we can infer from Fig.~\ref{fig:singlet}~(\emph{left}) an upper limit on the rate of a new physics decay mode of the $125\gev$ Higgs state. For instance, for a Higgs boson with identical couplings as in the SM, $\kappa=1$, we obtain $\mathrm{BR}(H\to \text{NP}) \lesssim 7.2\%$ (at $95\%~\mathrm{C.L.}$).\footnote{This limit is expected to improve to $ \mathrm{BR}(H\to \text{NP})  \le 4.3\%$ at the HL-LHC, see Sec.~6 of Ref.~\cite{Cepeda:2019klc}.}

Let's turn to the case where $h_S$ is heavier than $h_{125}$ and consider the collider signature (\emph{iii}). The maximal value of the signal rate $pp \to h_S \to h_{125} h_{125}$ at the $13~\mathrm{TeV}$ LHC that we can obtain in our model within all relevant theoretical and experimental constraints\footnote{This includes requirements of perturbative unitarity, boundedness of the potential, perturbative couplings as well as consistency with Higgs search limits, rate measurements and electroweak (EW) precision observables, see Refs.~\cite{Robens:2015gla,Robens:2016xkb,Ilnicka:2018def} for details. For the blue dotted line, we only impose these constraints at the EW-scale, whereas we apply them up to a high scale $\sim \mathcal{O}(10^{10}\gev)$ for the red solid line.} is shown in Fig.~\ref{fig:singlet}~(\emph{right}). We compare this with the current experimental limit from a combination of $h_S \to h_{125}h_{125}$ searches by CMS~\cite{Sirunyan:2018two}. We find that with current data, the experimental searches are not yet sensitive to this signature within the $Z_2$-symmetric real scalar singlet model. However, we expect that the LHC will become sensitive to parts of the parameter space with $h_S$ masses $\lesssim 500\gev$ in the high-luminosity phase. Complementary to $pp \to h_S \to h_{125} h_{125}$ searches are searches for the collider signature (\emph{ii}), most importantly, in the diboson final states, $pp\to h_S \to W^+ W^-$~and~$ZZ$. Current searches~\cite{Khachatryan:2015cwa,CMS:2017vpy,TheATLAScollaboration:2016bvt} already lead to constraints on the parameter space~\cite{Robens:2015gla,Robens:2016xkb,Ilnicka:2018def}.

\subsection{Adding two real scalar singlets}
\label{subsec:2singlets}

We now extend the SM scalar sector by two real scalar singlet fields, $S$ and $X$.\footnote{Here we show results from Ref.~\cite{Robens:2019kga}, where this model is studied in detail.} The scalar potential is then given by
\begin{align}
V(\Phi, S) =& + \mu_\Phi^2 \Phi^\dagger \Phi + \mu_S^2 S^2 + \mu_X^2 X^2 + \lambda_\Phi (\Phi^\dagger \Phi)^2 + \lambda_S S^4 + \lambda_X X^4\nonumber \\
&  + \lambda_{\Phi S} \Phi^\dagger \Phi S^2 + \lambda_{\Phi X} \Phi^\dagger \Phi X^2 + \lambda_{SX} S^2 X^2,
\end{align}
where we imposed a $\mathbb{Z}_2 \times \mathbb{Z}_2'$ discrete symmetry, with the following transformation properties:
$\mathbb{Z}_2:\, \Phi \to \Phi,\,S\to -S,\, X \to X,$ and $\mathbb{Z}_2':\, \Phi \to \Phi,\, S\to S,\, X \to -X$.
We focus here on the case that both singlet fields break the discrete symmetry spontaneously by acquiring a non-zero vev, $\langle S \rangle \ne 0$, $\langle X \rangle \ne 0$.\footnote{If one of the singlet fields has zero vev the corresponding scalar boson is stable and thus a DM candidate.} As a result, all three neutral scalar fields mix, forming the three physical scalar states $h_i$ ($i=1,2,3$) with masses $M_i$ (with $M_1 \le M_2 \le M_3$). As in the previous model in Sec.~\ref{subsec:singlet}, the couplings of the three scalar bosons to SM particles are again universally reduced by the mixing, whereas the Higgs self-couplings are determined by the scalar potential parameters and the vevs through the minimization conditions. For convenience, the model can be parametrized in terms of the three Higgs masses, $M_i$, the vevs $v=246\gev$, $\langle S \rangle$, $\langle X \rangle$, and the three rotation angles. One of the Higgs states $h_i$ has to be identified with the observed Higgs boson so that its mass is fixed by $M \simeq 125\gev$. We are left with seven free model parameters.

\begin{figure}[t]
\centering
\includegraphics[width=0.46\textwidth, trim=0cm 0.5cm 0cm 1cm]{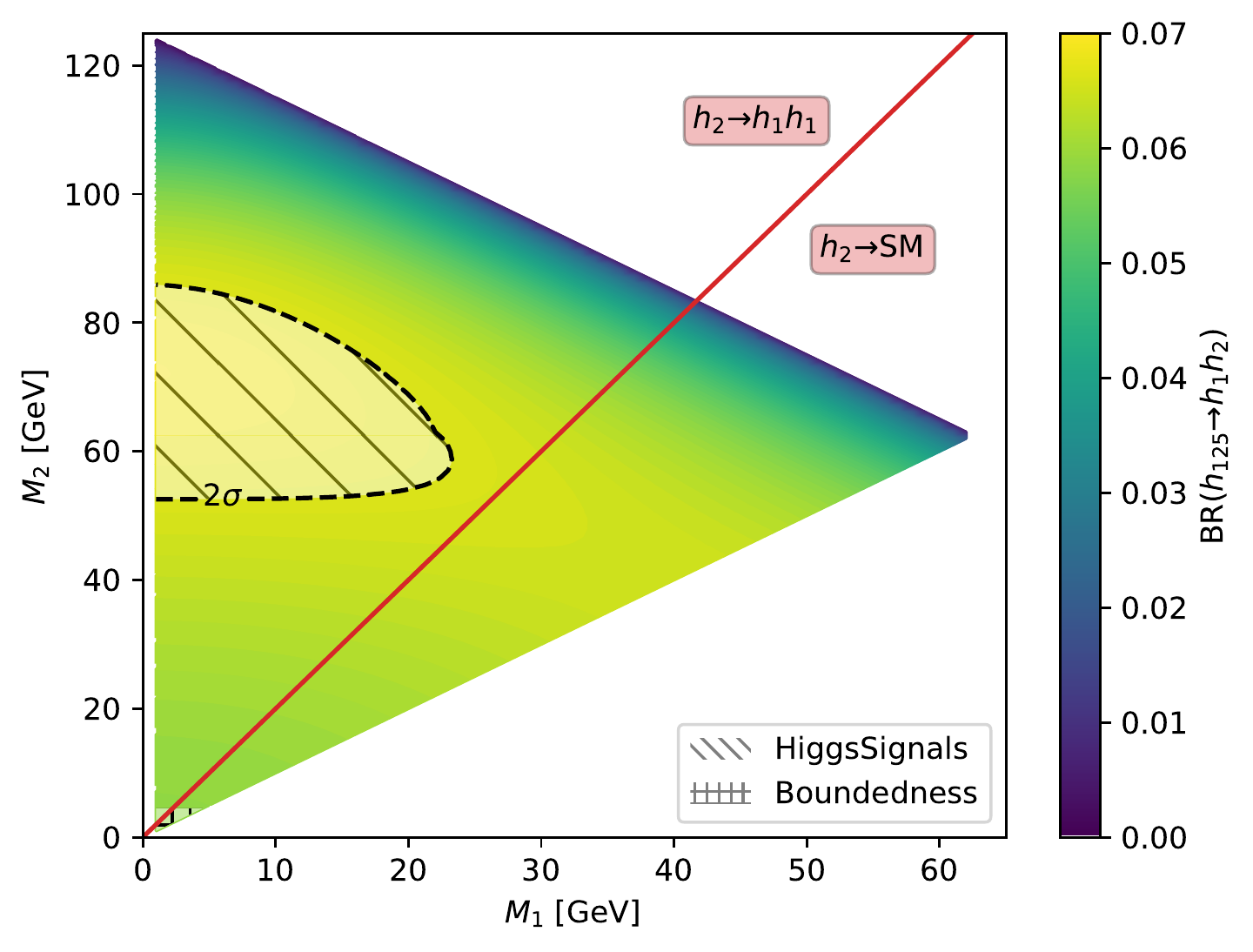} \hfill
\includegraphics[width=0.46\textwidth, trim=0cm 0.5cm 0cm 1cm]{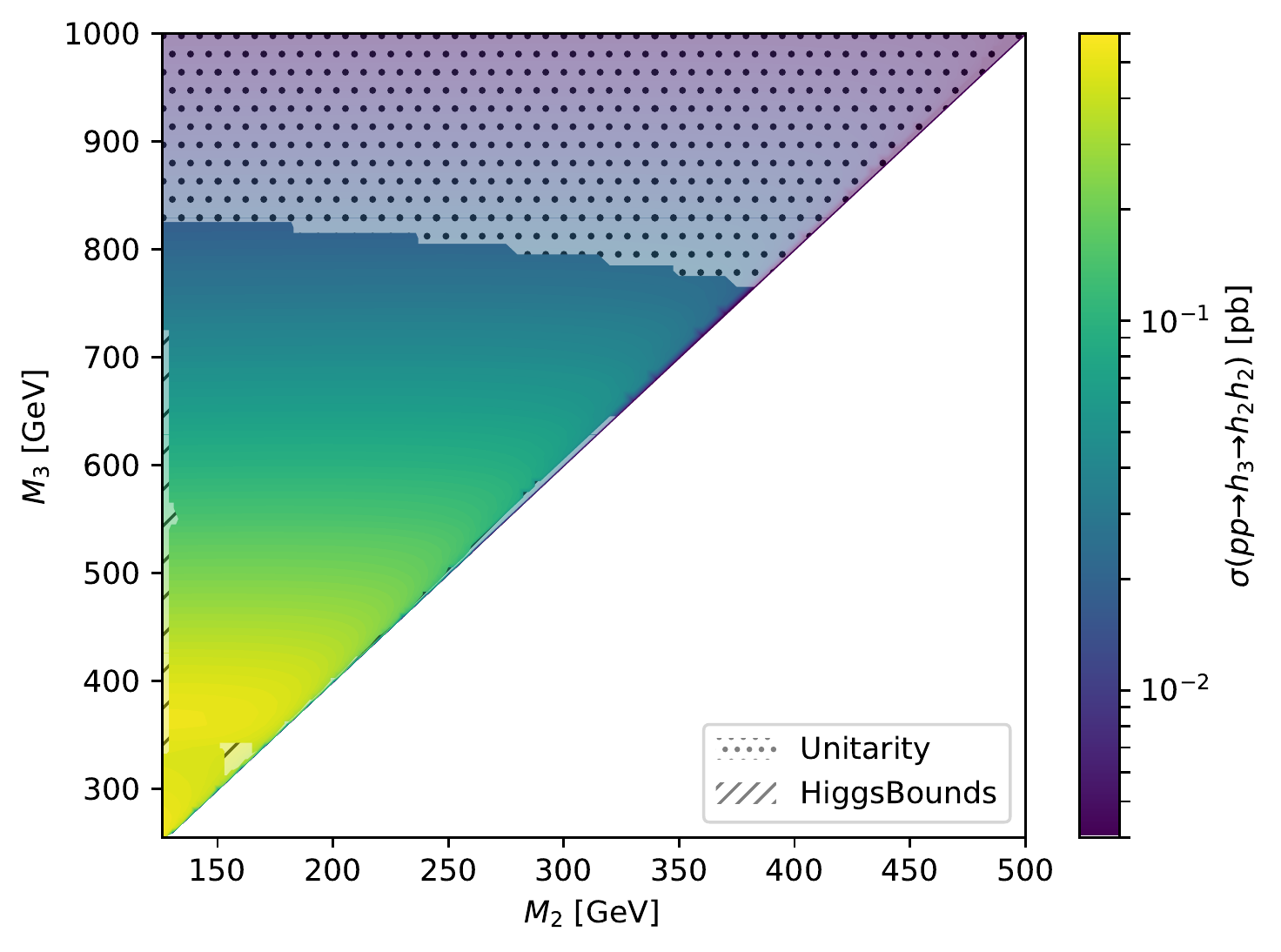}
\caption{\emph{Two real scalar singlet extension of the SM:} Benchmark planes \textbf{BP1} (\emph{left panel}) and \textbf{BP6} (\emph{right panel})  for Higgs-to-Higgs decay signatures at the LHC [taken from Ref.~\cite{Robens:2019kga}]. The hatched areas indicate excluded regions from theoretical or experimental constraints (\emph{see legend}). See text for further details.}
\label{fig:singlets}
\end{figure}

Regarding the collider phenomenology, the model features several interesting possibilities of Higgs-to-Higgs decay signatures. Fig.~\ref{fig:singlets} displays two representative benchmark planes (BP) (taken from Ref.~\cite{Robens:2019kga}) for LHC searches for Higgs-to-Higgs decays: In the first scenario, \textbf{BP1}, $h_3$ is identified with the observed Higgs boson at $125\gev$, and Fig.~\ref{fig:singlets}~(\emph{left}) shows the branching ratio (BR) for the decay $h_3 \to h_1 h_2$, amounting to $\sim (5 -7)\%$ in most of the parameter space. It is produced at nearly identical rates as the SM Higgs boson, i.e.~its total $13\tev$ LHC production rate is $\sim 50\pb$. The second lightest Higgs state, $h_2$, decays dominantly directly to SM particles (mostly $b\bar{b}$) if $M_2 < 2 M_1$ [\emph{below the red line}], or otherwise decays dominantly as $h_2\to h_1h_1$ [\emph{above the red line}], leading to a \emph{cascade} of Higgs-to-Higgs decays. The lightest scalar $h_1$ decays according to the SM Higgs prediction at its mass value $M_1$, i.e., mostly to $b\bar{b}$ if $M_1 \gtrsim 10\gev$. Therefore, in this benchmark scenario, the signal process is characterized by either two di-$b$-jet resonances at $M_1$ and $M_2$ in the invariant mass ($M_{bb}$) spectrum, or even three $M_{bb}$ resonances per event pointing to $M_1$. In the latter case where $h_2\to h_1h_1$, it may even be possible to reconstruct $M_2$ from four reconstructed b-jets. However, the experimental challenge of this signature is the softness of the final state objects, possibly demanding the presence of an associated particle in the production process (e.g., one additional jet, $pp \to h_3 + j$, or Higgs-Strahlung, $pp\to h_3 V$ [$V = W^\pm, Z$]).

In contrast, in the other benchmark plane, \textbf{BP6}, we assume $h_1$ to be at $125\gev$, and focus on the signature $pp\to h_3 \to h_2 h_2$. The $13\tev$ LHC signal rate is shown in Fig.~\ref{fig:singlets}~(\emph{right}). If $M_2 > 250\gev$, the decay $h_2\to h_1h_1$ happens to $\sim 30\%$, which in turn can lead to a spectacular cascade $pp\to h_3 \to h_2 h_2 \to h_1 h_1 h_1 h_1$, with a rate $\lesssim \mathcal{O}(10\fb)$. If $M_2 \le 250\gev$, the most promising signature is $pp\to h_3 \to h_2 h_2 \to W^+W^-W^+W^-$ and we find that a current ATLAS search~\cite{Aaboud:2018ksn} is already sensitive to small parts of the parameter space.



\section{Models with an additional scalar doublet}
\label{sec:doublets}

\subsection{The $CP$-conserving Two-Higgs-Doublet Model (2HDM)}
\label{subsec:2HDM}

We now extend the SM scalar sector by a second scalar $\mathrm{SU}(2)_L$ doublet field with hypercharge $+\tfrac{1}{2}$ (see Refs.~\cite{Gunion:1989we,Branco:2011iw} for reviews). With a softly-broken $\mathbb{Z}_2$ symmetry ($\Phi_1 \to  \Phi_1$, $\Phi_2 \to  -\Phi_2$), and assuming $CP$-conservation, the scalar potential in the \emph{general basis} reads
\begin{align}
V(\Phi_1,\Phi_2) = &+ m_{11} {\Phi_1}^\dagger \Phi_1 + m_{22} {\Phi_2}^\dagger \Phi_2 - [m_{12} {\Phi_1}^\dagger \Phi_2 + \mathrm{h.c.}] +\tfrac{1}{2} \lambda_1 ({\Phi_1}^\dagger \Phi_1)^2 +\tfrac{1}{2} \lambda_2 ({\Phi_2}^\dagger \Phi_2)^2 \nonumber \\
& + \lambda_3 ({\Phi_1}^\dagger \Phi_1)({\Phi_2}^\dagger \Phi_2) + \lambda_4 ({\Phi_1}^\dagger \Phi_2)({\Phi_2}^\dagger \Phi_1) + [\tfrac{1}{2} \lambda_5 ({\Phi_1}^\dagger \Phi_2)^2 +\mathrm{h.c.}]
\end{align}
with all parameters being real. The particle spectrum consists of two $CP$-even neutral Higgs bosons $h$ and $H$ (with masses $M_h \le M_H$), one $CP$-odd neutral Higgs boson $A$ (with mass $M_A$) and a pair of charged Higgs bosons $H^\pm$ (with mass $M_{H^\pm}$). In order to suppress dangerous flavor-changing neutral currents (FCNCs) at tree-level, the $\mathbb{Z}_2$ symmetry can be promoted to the fermion sector (with four different possible $\mathbb{Z}_2$ charge assignments to the fermion types). In a 2HDM of Type-I only $\Phi_2$ couples to the SM fermions, whereas in Type-II $\Phi_2$ couples to up-type quarks, and $\Phi_1$ couples to down-type quarks and leptons. 

In the general basis, the mixing of the two $CP$-even Higgs states is described by the rotation angle $\alpha$, and we define $\tan\beta \equiv v_2/ v_1$, where $v_{1,2}$ are the vevs of the neutral components of the doublet fields $\Phi_{1,2}$. The (SM normalized) Higgs couplings to vector bosons $V = W^\pm, Z$ for the physical $CP$-even Higgs states $h$ and $H$ are then given at tree-level by 
\begin{align}
\frac{g_{hVV}}{g_{h_\text{SM} V V}} = \sin (\beta - \alpha) \qquad \mbox{and} \qquad \frac{g_{HVV}}{g_{h_\text{SM} V V}} = \cos (\beta - \alpha).
\end{align}
In the two possible \emph{alignment limits}, $\sin(\beta - \alpha) \to 1$ or $\cos(\beta - \alpha) \to 1$, the Higgs state $h$ or $H$, respectively, has identical tree-level couplings to SM particles as predicted for a SM Higgs boson. Hence, either Higgs state $h$ or $H$ can be identified as the observed Higgs state at $125\gev$. It is therefore interesting to ask: \emph{Will we ever be able to distinguish these two cases?}

\begin{figure}[t]
\centering
\includegraphics[width=0.48\textwidth, trim=0cm 0.5cm 0cm 1cm]{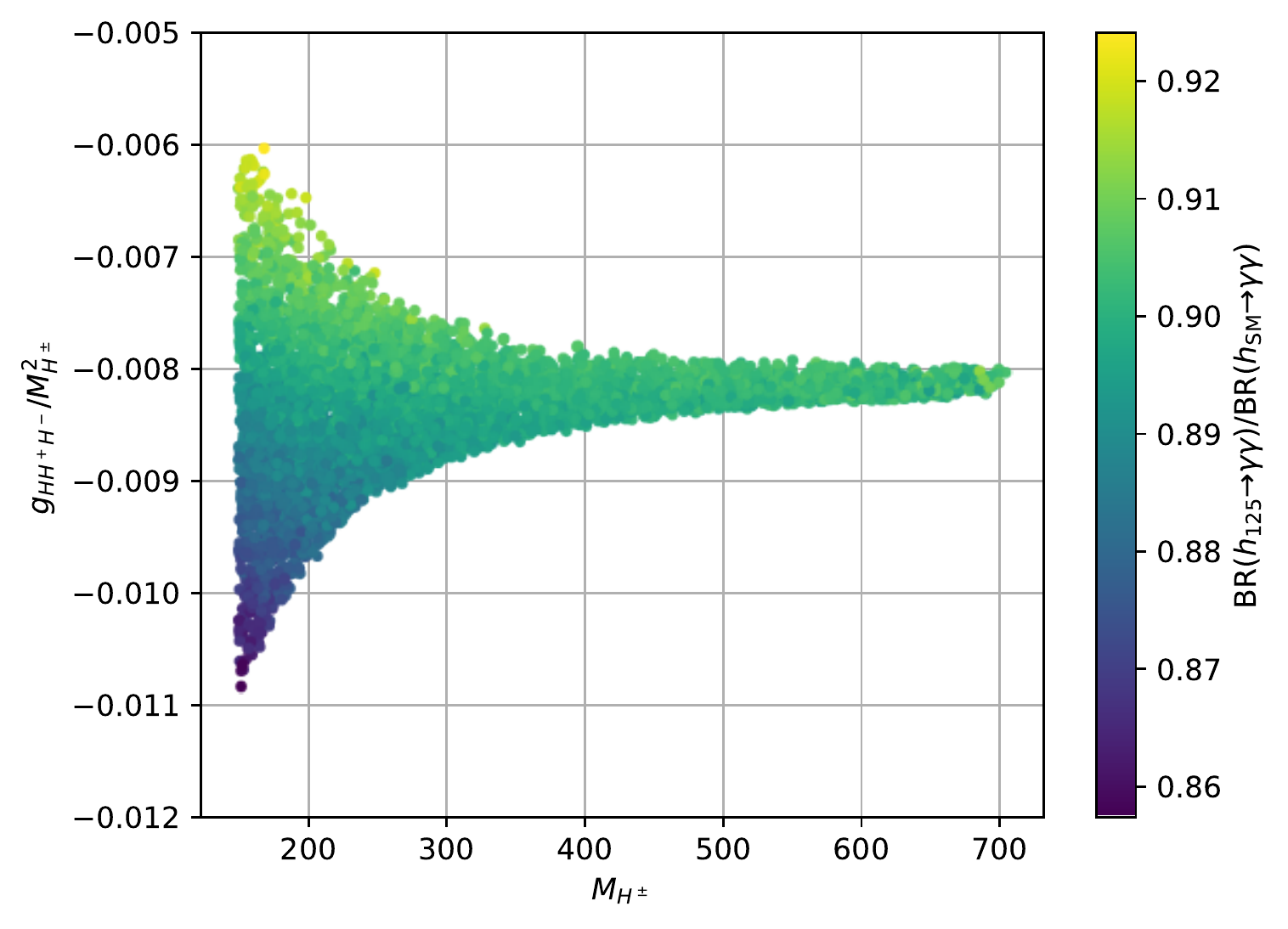}\hfill
\includegraphics[width=0.48\textwidth, trim=0cm 0.5cm 0cm 1cm]{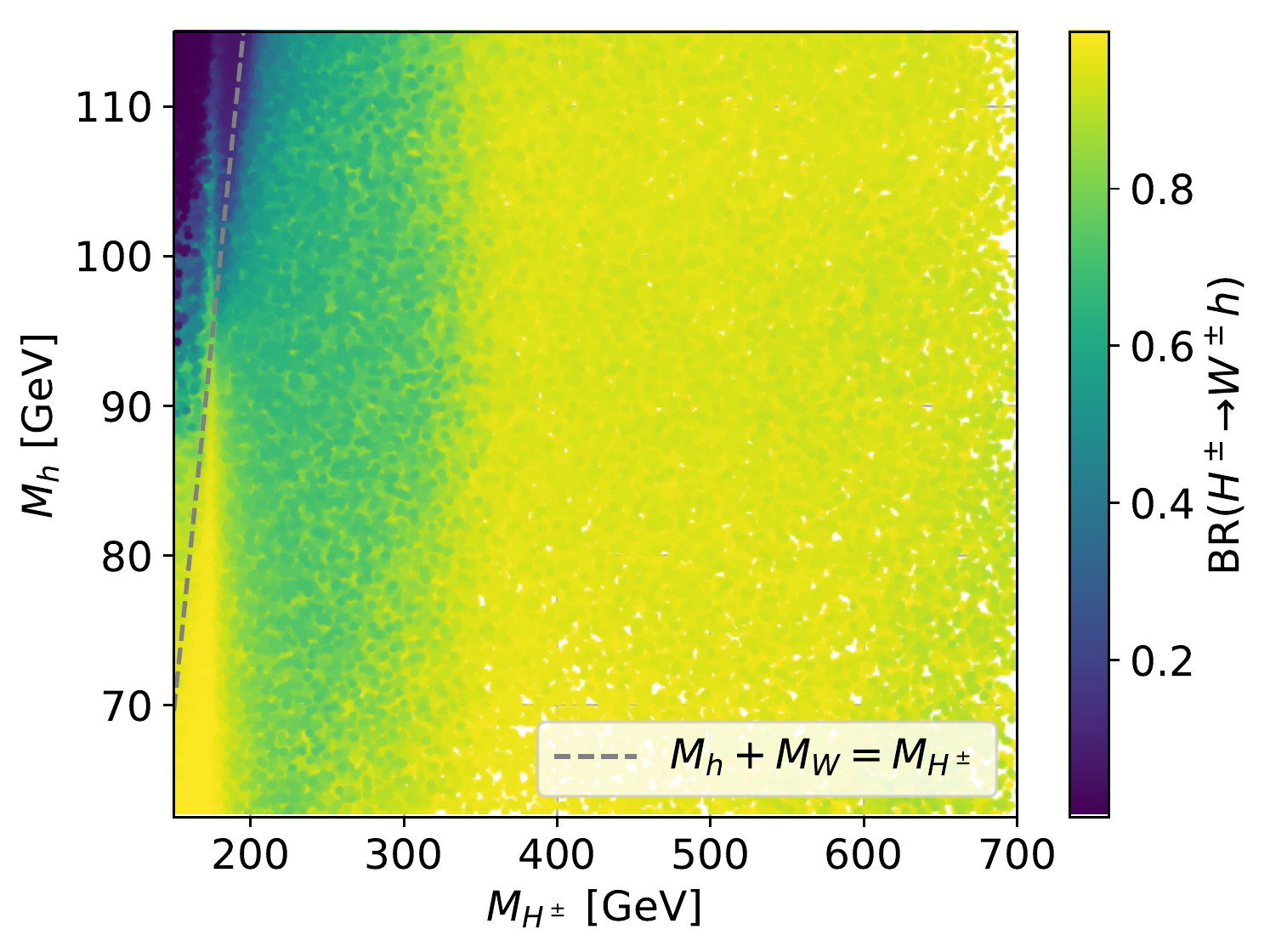}
\caption{\emph{Two-Higgs doublet model (2HDM) of Type-1}, with heavier Higgs boson $H$ at $125\gev$: Higgs-to-diphoton decay rate modification as a function of charged Higgs boson mass, $M_{H^\pm}$, and coupling $g_{HH^+ H^-}/M_{H^\pm}^2$ (\emph{left panel}); minimal decay rate $\mathrm{BR}(H^\pm \to W^\pm h)$ in the $(M_{H^\pm}, M_h)$ plane (\emph{right panel}).}
\label{fig:2HDM}
\end{figure}

It turns out that, within the 2HDM, the answer is yes, due to an interesting interplay between the neutral Higgs bosons $h, H$ (and possibly, $A$) with the charged Higgs boson, $H^\pm$. Let's assume the heavy Higgs state $H$ to be the observed Higgs boson. The Higgs rate measurements then require the alignment limit, $\cos(\beta-\alpha) \to 1$, to be approximate realized. Interestingly, contributions from $H^\pm$ to the $H\to \gamma \gamma$ decay neither decouple with large charged Higgs mass, $M_{H^\pm}$, nor vanish in the alignment limit~\cite{Bernon:2015wef}. The relevant coupling factor behaves as
\begin{align}
g_{HH^+H^-} \xrightarrow{c_{\beta-\alpha} \to 1} - \left( M_H^2  + 2 M_{H^+}^2 - 2\overline{m}^2 \right) / v  \xrightarrow{M_{H^+} \gg M_H} - 2M_{H^+}^2/v,
\end{align}
because $\overline{m}^2 \equiv 2m_{12}^2/\sin(2\beta) \lesssim \mathcal{O}(v^2) $ imposed by unitarity and stability conditions. It follows that, if the charged Higgs boson is heavy, it will leave an observable trace in the $H\to \gamma\gamma$ rate. This is illustrated in Fig.~\ref{fig:2HDM} (\emph{left}) for the 2HDM of Type-1, showing the rate modification $\mathrm{BR}(H\to \gamma\gamma)/\mathrm{BR}(h_\text{SM}\to \gamma\gamma)$ as a function of $M_{H^\pm}$ and $g_{HH^+H^-}/M_{H^\pm}^2$ for all allowed parameter points with $H$ at $125\gev$. We find a decay rate modification of around $-10\%$ at large $M_{H^\pm}$.

This brings us to the question: How light can the charged Higgs boson be? In the 2HDM of Type-2 flavor observables --- in particular the $B \to X_s \gamma$ decay rate --- severely constrain the charged Higgs mass, $M_{H^\pm} \gtrsim 600\gev$~\cite{Arbey:2017gmh}. In contrast, in Type-1 $M_{H^{\pm}}$ is essentially unconstrained by flavor observables if $\tan\beta\gtrsim 2$. Here, LHC searches for a light (or moderately heavy) charged Higgs boson will be crucial. Indeed, as the second $CP$-even Higgs boson $h$ is very light, $M_h < M_H = 125\gev$, and the coupling $g_{H^\pm W^\mp h} \propto \cos(\beta-\alpha)$ is maximal in the alignment limit, the decay $H^\pm \to W^\pm h$ is generally dominant. The \emph{minimal} value of its decay rate for given values of $M_h$ and $M_{H^\pm}$ is shown in Fig.~\ref{fig:2HDM} (\emph{right}) for the allowed Type-1 parameter points.\footnote{The maximal $\text{BR}(H^\pm \to W^\pm h)$ value is close to $100\%$ in almost the whole mass plane.} 
Most of the current $H^\pm$ searches at the LHC, however, focus on the fermionic final states ($\tau \nu_\tau$, tb), which are insensitive to these scenarios. Direct searches for $H^\pm \to W^\pm h$ decay signatures will therefore be crucial to conclusively discriminate between the $h$ and $H$ interpretation of the observed Higgs state~\cite{Stefaniak:inprogress}.



\subsection{The Minimal Supersymmetric Standard Model (MSSM)}
\label{subsec:MSSM}

\begin{figure}[t]
\centering
\includegraphics[width=0.46\textwidth, trim=0cm 0.5cm 0cm 1cm]{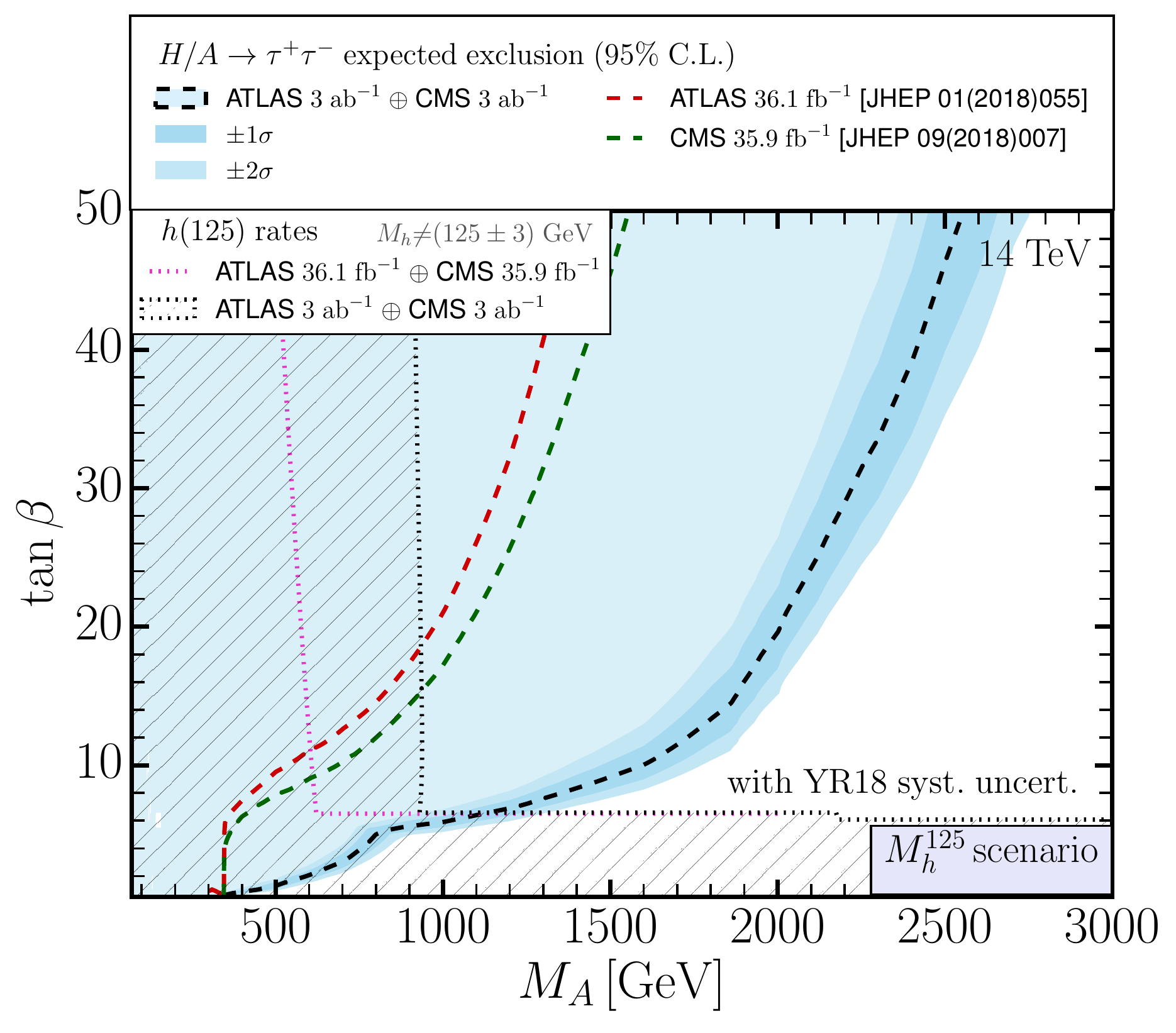}\hfill
\includegraphics[width=0.46\textwidth, trim=0cm 0.5cm 0cm 1cm]{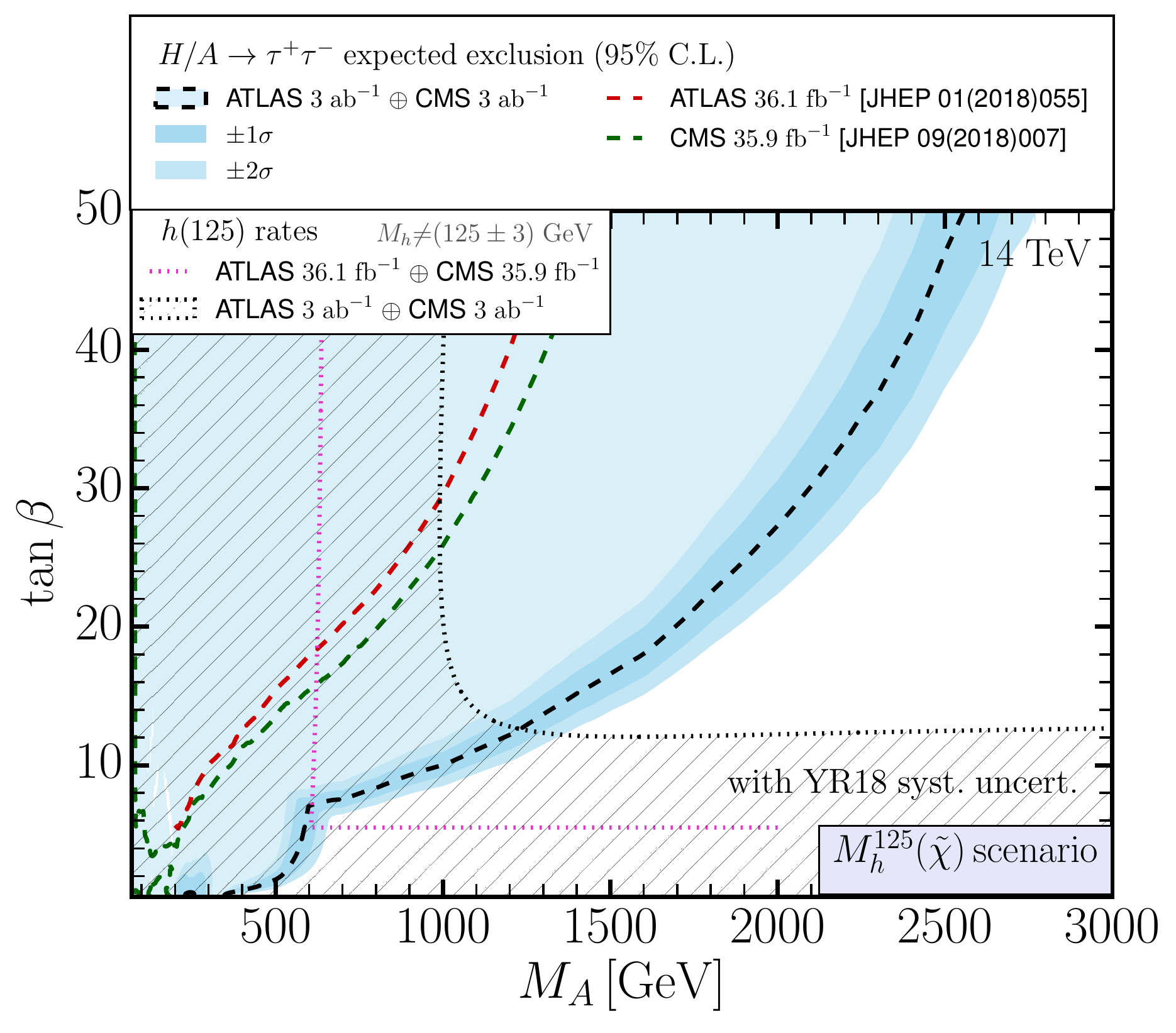}
\caption{HL-LHC prospects for the MSSM Higgs sector, presented in the $M_h^{125}$ (\emph{left panel}) and $M_h^{125}(\tilde\chi)$ scenario (\emph{right panel}), taken from Sec.~9.5 of Ref.~\cite{Cepeda:2019klc}.}
\label{fig:MSSM_HL-LHC}
\end{figure}

At tree-level, the MSSM Higgs sector is a 2HDM Type-2 with quartic couplings fixed by the gauge couplings. It can threrefore be described in terms of two parameters, often chosen to be $M_A$ and $\tan\beta$. However, beyond tree-level, all SUSY parameters affect the Higgs sector. Besides precision Higgs mass and rate measurements, LHC searches for the heavier neutral Higgs bosons $H$ and $A$ decaying to $\tau^+\tau^-$ probe sensitively the parameter space. In Fig.~\ref{fig:MSSM_HL-LHC} we show the current and future HL-LHC sensitivity to the MSSM Higgs sector, employing the recently proposed $M_h^{125}$ and $M_h^{125}(\tilde\chi)$ benchmark scenario~\cite{Bahl:2018zmf}, via Higgs signal rate measurements and $H/A\to \tau^+\tau^-$ searches~\cite{Cepeda:2019klc}. Heavy Higgs masses below $1\tev$ will be completely probed. In the $M_h^{125}(\tilde\chi)$ scenario the $H/A\to \tau^+\tau^-$ reach is weakened due to additional $H/A$ decay modes to light neutralinos and charginos, $H/A \to \tilde{\chi}\tilde{\chi}$. Dedicated experimental searches for these decays would be highly complementary and may improve the coverage in the moderate $\tan\beta$ region.

\section{Conclusions}
\label{sec:conclusions}

We discussed the phenomenological status of simple and popular BSM Higgs sectors, including scalar singlet extensions of the SM, the 2HDM and the MSSM Higgs sector. The LHC results on the $125\gev$ Higgs boson and searches for additional Higgs states have important implications for BSM Higgs models, and imply that an approximate \emph{alignment limit} (i.e.~SM-like Higgs couplings at tree-level) is realized. Nevertheless, there is still room for new Higgs discoveries in upcoming LHC runs. Additional Higgs states can be lighter or heavier than the discovered Higgs boson, and experimental searches should aim to cover the full accessible kinematical range. Furthermore, some LHC searches only become sensitive with more data, as illustrated here for LHC searches for resonant double Higgs production in the $\mathbb{Z}_2$-symmetric singlet extension(s). We also pointed out so-far-uncovered collider signatures, including Higgs-to-Higgs decays ($h_i \to h_j h_k$ and $H^\pm \to W^\pm h$), and heavy Higgs decays to neutralinos and charginos ($H/A \to \tilde{\chi}\tilde{\chi}$).

\section*{Acknowledgments}

We thank the ALPS 2019 organizers for a very stimulating workshop and their hospitality. We are grateful to Jonas Wittbrodt for many helpful discussions.  This work is funded by the Deutsche Forschungsgemeinschaft (DFG, German Research Foundation) under Germany‘s Excellence Strategy -- EXC 2121 ``Quantum Universe'' -- 390833306.

\end{document}